\documentstyle[preprint,aps,psfig]{revtex}

\begin{document}

\tighten
\draft
\preprint{
\vbox{
\hbox{May 1995}
\hbox{TUM/T39-95-7}
\hbox{ADP--95--29/T183}
}}

\title{Meson Cloud of the Nucleon in
       Polarized Semi-Inclusive Deep-Inelastic Scattering}
\author{W.Melnitchouk}
\address{Physik Department,
         Technische Universit\"{a}t M\"unchen,
         D-85747 Garching, Germany.}
\author{A.W.Thomas}
\address{Department of Physics and Mathematical Physics,
         University of Adelaide,
         5005, Australia.}

\maketitle

\begin{abstract}
We investigate the possibility of identifying an explicit pionic
component of the nucleon through measurements of polarized
$\Delta^{++}$ baryon fragments produced in deep-inelastic
leptoproduction off polarized protons, which may help to identify
the physical mechanism responsible for the breaking of the Gottfried
sum rule.
The pion-exchange model predicts highly correlated polarizations of
the $\Delta^{++}$ and target proton, in marked contrast with the
competing diquark fragmentation process.
Measurement of asymmetries in polarized $\Lambda$ production may
also reveal the presence of a kaon cloud in the nucleon.
\end{abstract}
\pacs{PACS numbers: 13.60.Hb, 13.87.Fh, 13.88.+e   \vspace*{2cm}\\
      To appear in {\em Zeit.Phys.A}}

\section{INTRODUCTION}

The remarkably successful application of the quark--parton model
in the description of deep-inelastic scattering (DIS) data over
a very large kinematic domain has propelled this
simple picture of the nucleon at high energies into becoming part
of the common language employed by medium and high energy physicists.
Furthermore, the QCD-improved parton model provides a framework
in which one can quantitatively understand the scaling violations
seen in DIS experiments in the perturbative region of large photon
virtualities ($Q^2 \agt 4$ GeV$^2$).
Nevertheless, what we have learned from the more recent DIS data
on both polarized and unpolarized targets is that even the QCD-improved
parton model cannot, by itself, give a complete description of the
structure of the nucleon at high energies.
It is unable to (nor was it intended to) explain the spectrum
of the nucleon's non-perturbative features.
Here one has traditionally invoked effective degrees of freedom,
for example in the form of a pionic cloud of the nucleon, to
describe the long range structure of the nucleon.

A very good example of this is the deviation from the QCD-parton
model prediction for the Gottfried sum rule \cite{GSR} seen in
the recent high-precision NMC data \cite{NMC}.
The most natural explanation of this result is that there exists
an excess of $\bar d$ quarks over $\bar u$ in the proton
--- something which is clearly impossible to obtain from
perturbative QCD alone.
A non-perturbative pionic cloud, on the other hand, offers
a simple explanation of this SU(2) flavor symmetry breaking
in the proton sea \cite{HM,SST,MTS,KL,MA,JUL}.
The more recent NA51 Drell-Yan experiment \cite{NA51} also strongly
suggests a suppression of the $\bar u$ sea in the proton
relative to the $\bar d$ sea.

Similarly in polarized DIS, the small value for the first moment
of the proton's spin-dependent structure function, $g_1$, obtained
initially by the EMC \cite{EMCG1}, and confirmed by later
measurements at CERN \cite{SMC} and SLAC \cite{SLACG1},
is widely interpreted as evidence of the breakdown of the simple
quark--parton model of nucleon structure.
The two most common interpretations of this result are
that either the strange sea of the proton is significantly
polarized, or that subtle anomaly effects (perhaps in the form
of highly polarized gluons) lead to a strong violation
of the OZI rule in the flavor singlet channel \cite{U1GLU}.
The simplest way to model a polarized strange sea would be in terms
of a polarized hyperon accompanying a non-perturbative cloud of kaons
\cite{TTT}.
DIS from a $\Lambda K$ or $\Sigma K$ component of the nucleon would
be a natural mechanism leading to a violation of the Ellis-Jaffe sum
rule \cite{EJ}.

While strongly suggesting that non-perturbative effects play an
important role in nucleon DIS, the results of these experiments
do not rule out mechanisms other than those involving meson clouds
as those responsible for the deviations of these sum rules from
the parton model predictions.
Indeed, despite the various phenomenological successes of nucleon
models which incorporate mesonic degrees of freedom, as yet there
is no direct experimental evidence unambiguously pointing to the
existence of a pion (or kaon) cloud in high energy reactions.
It is the purpose of this paper to identify experiments which could
give clear and unique signals of the presence of mesonic degrees
of freedom in nucleon DIS.

The role of pions in inclusive DIS from nucleon has been investigated
in a number of previous studies \cite{HM,SST,MTS,KL,MA,JUL,SUL,T83,FMS}.
Following Thomas \cite{T83}, it was realized that upper bounds
on the average pion number per nucleon could be extracted by
comparing with DIS data on the momentum fractions carried by
sea quarks in the proton.
Controversy as to whether all or just part of the Gottfried sum
rule violation can be accounted for by the pion cloud could be resolved
by obtaining a {\em lower} bound on the pion multiplicity.
However, to obtain a lower bound one would need to extract a
clear pionic signal from beneath the background arising from
the perturbative sea.
Since the pion contribution to the nucleon structure function
appears at relatively small Bjorken $x$ ($x \sim 0.1$), its signal
may be submerged beneath the perturbative background.
Therefore it seems a formidable challenge to seek direct
experimental confirmation of pionic effects in inclusive
DIS.
The problem is worse for the case of the $K$ cloud since, being
heavier, those contributions lie at even smaller $x$.

The pertinent question to ask is whether pions leave any unique
traces at all in other processes, which cannot be understood in
terms of perturbative quark and gluon degrees of freedom alone.
Recently in the literature several suggestions have been made
regarding the measurement of the pion cloud in other experiments.
Pirner and Povh \cite{HEID} have proposed to identify the size of
the constituent quark--pion vertex through exclusive leptoproduction
of fast pions in the current fragmentation region.
Dieperink and Pollock \cite{DP} have argued that one could obtain
information on the $\pi N$ form factor in DIS from a $^3He$ nucleus
by detecting the recoiling $^3He$ nucleus in the final state.
In the present paper we propose a series of analogous experiments
in semi-inclusive DIS on polarized protons, where a hadron is detected
in the final state in coincidence with the
scattered electron \cite{STA,LUS}.
We will demonstrate that DIS from the nucleon's pion cloud
(Fig.1) does in fact give rise to rather characteristic
fragmentation distributions in comparison with the predictions
of parton model hadronization.
These differences are significantly enhanced when initial
and final state polarization effects are considered.

We focus on semi-inclusive production of polarized $\Delta^{++}$
baryons from a polarized proton,
$e \vec p \rightarrow e' \vec\Delta^{++} X^-$.
Because the $g_1$ structure function of a pion is zero,
an unpolarized electron beam will suffice for this purpose.
The choice of the $\Delta^{++}$ for the final state baryon,
rather than, say, a nucleon, reduces the backgrounds that one
would have to consider due to the decay of $\Delta$s themselves.
Furthermore, the decay products of $\Delta^+$ or $\Delta^0$
would include neutral hadrons whose detection would be more
difficult, thus increasing the overall experimental uncertainties.
For the case of the $K$ cloud, the relevant reaction to observe
is $e \vec p \rightarrow e' \vec \Lambda X^+$.
Determining the polarization of the $\Lambda$ hyperon is
considerably easier because the $\Lambda$ is self-analyzing.

In the next Section we outline the basic kinematics
pertinent to semi-inclusive deep-inelastic scattering.
In Section III we present the predictions for the
polarization asymmetries in the pion cloud model
of the nucleon.
A condensed summary of the results in this Section can also
be found in Ref.\cite{CEBAF}.
Possible backgrounds to the pionic signal are
analyzed in Section IV.
In Section V the strangeness content of the nucleon is
studied in the $K$ cloud and diquark fragmentation models.
A brief overview of other experiments
suggested recently to measure the pion cloud of the nucleon
is given in Section VI, while Section VII is reserved for
some concluding remarks.

\section{KINEMATICS OF TARGET FRAGMENTATION}

Experimentally it is known that the yield of baryons is about
one order of magnitude higher in the backward hemisphere of the
$\gamma p$ center of mass frame (``target fragmentation region'')
than for forward hemisphere baryons (``current fragmentation region'')
\cite{BEB77,ARN85,EMC86}.
Furthermore, baryons produced by current fragmentation have
predominantly large momenta in the target rest frame
($\agt$ several GeV), while those in the backward center
of mass jet are generally slow.
Since our concern here is with low momentum baryons ($\Delta$s and
$\Lambda$s) produced in the target fragmentation region, we shall
neglect the quark $\rightarrow$ baryon fragmentation process which
gives rise to the forward baryons.

For studies of the spin dependence of the fragmentation process,
we require the target proton polarization to be parallel to the
photon direction, with the spin of the produced baryon quantized
along its direction of motion.
Experimentally, the polarization of the produced $\vec\Delta^{++}$
can be reconstructed from the angular distribution of its decay
products ($p$ and $\pi^+$), while because it is self-analyzing,
polarization of the $\Lambda$ can be determined automatically.

With the high luminosity beam available at CEBAF, for example,
the rate of $\Delta^{++}$ (or $\Lambda$) production will generally
be high.
Even though the efficiency with which low momentum baryons
can be accurately identified is lower than for fast baryons
in the forward center of mass hemisphere \cite{EMC86}, their
detection will still be feasible, for example with the CEBAF
Large Acceptance Spectrometer.
Alternatively, with polarized internal targets soon available
at HERMES, one could also in principle perform this experiment
there provided the $4\pi$ detectors will be capable of
capturing slow moving baryons as well as the fast mesons,
which will be the focus of the first stage of the HERMES
program.
Such a program would also be ideally suited for the proposed
European electron facility, ELFE, or the new Hadron Muon Collaboration
at CERN, which will be designed specifically for semi-inclusive
measurements.

We define our variables in the target rest frame as follows:
$l, l'$ are the four-momentum vectors of the initial and final
leptons;
$P_{\mu} = (M; 0, 0, 0)$ and
$p_{\mu} = (p_{0}; |{\bf p}| \sin\alpha \cos\phi,$
          $|{\bf p}| \sin\alpha \sin\phi,$
          $|{\bf p}| \cos\alpha)$
are the momentum vectors of the target proton and recoil baryon,
respectively; and
$q_{\mu} = (\nu; 0, 0, \sqrt{\nu^{2} + Q^{2}})$ denotes the photon
four-momentum, defined to lie along the positive $z$-axis.
Then $\nu = E-E'$ is the energy transferred to the target,
$y = \nu / E = 1 - E'/E$ is the fractional energy transfer
relative to the incident energy, and $Q^2 = -q^2 = 2 M E x y$\
is minus the four-momentum squared of the virtual photon, with
$x = Q^2 / 2 P \cdot q$.
With the possible CEBAF upgrade to $E \approx 8$--10 GeV, values of
$x \approx 0.13$--0.14 can be reached in the deep-inelastic region
for $\nu \approx 8$ GeV and $Q^2 \approx 2$ GeV$^2$,
corresponding to a center of mass energy squared of the
$\gamma p$ system of $W^2 = (P + q)^2 \sim 15$ GeV$^2$.
At HERMES, with a 30 GeV electron beam, one will comfortably probe
the $0.05 \alt x \alt 0.1$ region,
which is relevant for the pionic contribution,
up to $Q^2 \sim 5$ GeV$^2$ and $W^2 \alt 50$ GeV$^2$.

The four-momentum transfer squared between the proton and
baryon is
$t \equiv (P-p)^2 = -p^2_T / \zeta\ +\ t_{max}$,\
which is bounded from above
by $t_{max} = -(M_{B}^2 - M^2 \zeta) (1-\zeta)/\zeta$,
where $p_T^2 = {\bf p}^2 \sin^2\alpha$,
$\zeta = p \cdot q / P \cdot q$\ \ is the light-cone fraction
of the target proton's momentum carried by the baryon,
and $M_B$ is the recoiling baryon's mass.
In terms of $t$, the three-momentum of the produced baryon
is given by:
\begin{eqnarray}
|{\bf p}| &=& \frac{1}{2M} \sqrt{(M^2 + M_{B}^2 - t)^2
                                      - 4 M^2 M_{B}^2}\ ,
\end{eqnarray}
so that in the target rest frame the slowest baryons are those
for which $t$ is maximized,
which occurs when $\zeta \rightarrow 1$.
As the upper limit on $\zeta$ is $1-x$, slow
baryon production also corresponds to the $x \rightarrow 0$ limit,
and the slowest possible particles produced at $\zeta = 1$
(at $x = 0$) will have momentum
$|{\bf p}_{min}| = (M_{B}^2 - M^2) / 2M \approx 340$ MeV
for $B=\Delta$,
and $\approx 193$ MeV for $B=\Lambda$.
For the pion-exchange process considered here, the peak in the
differential cross section occurs at $|{\bf p}| \sim 600$ MeV,
which, for $\zeta \sim 0.8$, corresponds to a missing mass of
$p_X^2 = (P-p+q)^2 \sim 0.8$ GeV$^2$
for $Q^2 \sim 2$ GeV$^2$ at CEBAF energies,
and
$p_X^2 \sim 5$ GeV$^2$
for $Q^2 \sim 4$ GeV$^2$ at HERMES.

In terms of the polar angle $\alpha$ (in the target rest frame),
\begin{eqnarray}
\cos\alpha
&=& \frac{ M_{B}^{2} + (1-2 \zeta) M^{2} - t }
         { \sqrt{(M_{B}^{2} - M^{2} - t)^{2} - 4 M^{2} t} },
\label{calpha}
\end{eqnarray}
between the $B$ and $\gamma$ momenta,
production of baryons will occur between $\alpha = 0$ and
\begin{eqnarray}
\alpha_{max}
&=& \arccos \left( \sqrt{1 - (M \zeta / M_{B})^2}
              \right),
\end{eqnarray}
which for $\zeta \rightarrow 1$ is $\simeq 50^o$ for $B=\Delta$
and $\simeq 57^o$ for $B=\Lambda$.

For a given angle $\alpha$, the pion four-momentum will be
constrained to lie within the limits given by:
\begin{eqnarray}
t_{min/max}(\alpha)
&=& {1 \over \sin^2\alpha}
    \left( M_{B}^2 \sin^2\alpha
         - M^2 (1 - 2 x + \cos^2\alpha) \right.
\nonumber\\
&\pm&
    \left. 2 M \cos\alpha\ \
           \sqrt{ M^2 (1-x)^2
         - M_{B}^2 \sin^2\alpha} \right).
\end{eqnarray}
At small angles baryons will be produced over essentially
the entire range of $t$
(and therefore $\zeta$), however the number will fall off rapidly as
$\alpha \rightarrow \arccos \left( \frac{1}{M_{B}}
                           \sqrt{M_{B}^2 - M^2 (1-x^2)}
                    \right)$
because of the fast convergence of the upper and lower bounds on $t$,
until no particles are produced beyond the kinematic boundary at
$t_{max} = t_{min}$
$ = - \left( M_{B}^2 (1 + x) - M^2 (1 - x) \right) / (1-x)$.

The importance of the above kinematic limits
was demonstrated in two experiments \cite{E745,BEBC}
in which slow proton production was studied in $\nu$-nucleon
and $\nu$-nucleus scattering.
The softening of the cross section for protons with
momentum less than ${\bf p}_{max}$ (equal to several hundred MeV in
the experiments), was shown \cite{IST,GRO,MTN} to be precisely due
to the absence of interactions at $x > x_{max}$, where\
$x_{max}
= 1 - (p_{0max} - |{\bf p}_{max}|)/M$.

The role of pions was also investigated in this process,
however due to the large perturbative sea component of the
nucleon structure function at $x \sim 0.05$, no definite
pionic signal could be identified.
We may hope, however, that by including polarization degrees
of freedom we can more efficiently isolate any pionic signal from
behind the fragmentation background.

\section{PION CLOUD DYNAMICS}

The pion model is a dynamical model of the nucleon where the
dissociation of a physical nucleon into a pion and an ``undressed''
nucleon or $\Delta$ is explicitly witnessed by the probing photon.
The possible relevance of the process illustrated in Fig.1, where
a $\pi^-$ emitted by the proton is hit by a photon, to DIS was
recognized some time ago \cite{SUL,STA,LUS}, and has since had several
important and interesting applications, most notably in providing
a mechanism to break SU(2) and SU(3) flavor symmetries in the
proton sea.
In the pion-exchange model the differential cross section is:
\begin{eqnarray}
{ d^5 \sigma \over dx dQ^2 d\zeta dp_T^2 d\phi }
&\propto&
    { f_{\pi N \Delta}^2 \over 16 \pi^2 m_{\pi}^2 }\
    { {\cal T}^{S\ s}(t)\ {\cal F}^2_{\pi \Delta}
      \over (t - m_{\pi}^2)^2 }\
    L_{\mu\nu}(l,q)\ W_{\pi}^{\mu\nu}(k,q),
\label{ope5}
\end{eqnarray}
where
$L_{\mu\nu} = 2 l'_{\mu} l_{\nu} + 2 l'_{\nu} l_{\mu}
            - g_{\mu\nu} Q^2$
is the lepton tensor,
and
\begin{eqnarray}
W_{\pi}^{\mu\nu}
&=& - \left( g^{\mu\nu} + \frac{ q^{\mu} q^{\nu} }{ Q^2 } \right)\
             W_{1\pi}\
 +\  \left( k^{\mu} + \frac{ k \cdot q }{ Q^2 } q^{\mu} \right)
     \left( k^{\nu} + \frac{ k \cdot q }{ Q^2 } q^{\nu} \right)\
     \frac{ W_{2\pi} }{ m_{\pi}^{2} },
\label{Wpi}
\end{eqnarray}
describes the $\gamma \pi$ vertex, with $k$ denoting the virtual
pion four-momentum.
The quantity $T^{S\ s}(t)$ is the amplitude for a nucleon of
spin $S$ to emit a pion of four-momentum squared $t$, leaving
a $\Delta$ with spin $s$.

Since in the final analysis we will be dealing with Lorentz-invariant
cross sections as a function of the Lorentz-scalars $x$ and $\zeta$,
we can, without loss of generality, formulate the problem in any
frame which will simplify the analysis.
Here we note that factorization of the $\gamma N$ cross section
into $\gamma\ \pi$ and $\pi\ N$ (or $\gamma\ N^*$ and $N^* N$)
cross sections does not hold in all frames of reference \cite{MST}.
Indeed, such factorization, or convolution, can only be achieved
by eliminating antiparticle degrees of freedom, which can
formally be done only in the infinite momentum frame (IMF)
or on the light-cone \cite{WEIN,DLY}.
Therefore for the $\pi N \Delta$ form factor in Eq.(\ref{ope5})
we take the form suggested in earlier IMF studies of the
pionic content of the proton in inclusive DIS \cite{MTV}:
\begin{mathletters}
\begin{eqnarray}
\label{FF}
{\cal F}_{\pi \Delta}(p_T^2,\zeta) &=&
\left( { \Lambda^2 + M^2 \over
         \Lambda^2 + s_{\pi \Delta} }
\right)^2,
\end{eqnarray}
where $s_{\pi \Delta} \equiv (p + k)^2
                      = (m_{\pi}^2 + p_T^2)/(1-\zeta)
                      + (M_{\Delta}^2 + p_T^2)/\zeta$.
Since the form factor in the IMF is not yet very well
constrained, other forms for its shape \cite{ZOL} are also
possible (it has been suggested in Ref.\cite{ZOL} to use
semi-inclusive $NN$ scattering data as a means of obtaining
an upper bound on $\Lambda$, although here one also has to
deal with contributions from competing Reggeized meson
exchanges \cite{AG}).
However, the precise shape of the form factor is not
important here, since, as we shall see, the bulk of the effect
is given entirely by the proton--pion spin correlations.
Indeed, covariant formulations with $t$-dependent form factors
\cite{HM,SST,MTS,KL,MA,JUL,MSM}:
\begin{eqnarray}
\label{FFt}
{\cal F}_{\pi \Delta}(p_T^2,\zeta) &=&
\left( { \Lambda^2 - M^2 \over
         \Lambda^2 - t(p_T^2,\zeta) }
\right)^2,
\end{eqnarray}
\end{mathletters}
give very similar results to those with the
$s_{\pi\Delta}$-dependent forms in Eq.(\ref{FF}).

The formulation in the IMF also allows one to use the on-mass-shell
structure function of the pion in Eq.(\ref{Wpi}) \cite{MTV,ZOL,GCPP},
without the need to model the extrapolation of the off-shell pion
structure function into the $t \not= m_{\pi}^2$ region
\cite{PISF,SHAKIN}.
(Although, in principle, there could be effects in the virtual pion
structure function due to the off-energy-shell dependence.)
For the pion structure function we use therefore the most
recent parametrization \cite{SMRS} of data extracted from
Drell-Yan experiments \cite{NA10}.
The main uncertainty in the covariant calculation is in fact the
off-mass-shell extrapolation of the virtual pion structure function,
for which there still does not exist consensus in the literature
\cite{PISF,SHAKIN}.

The $\pi N \Delta$ coupling constant, $f_{\pi N \Delta}$, in
Eq.(\ref{ope5}) is the physical coupling constant, defined
at the pion pole ($t = m_{\pi}^2$).
Note that there is no renormalization factor, $Z$,
multiplying the pion-exchange cross section, as has been
used recently in Refs.\cite{JUL,SB}.
This factor, which to first order in $f_{\pi N \Delta}$ is
written $Z = 1 / (1 + < n >_{\pi\Delta})$, with the pion number
$< n >_{\pi\Delta}$ being essentially the integrated cross section
in Eq.(\ref{ope5}), is usually introduced to normalize the total
nucleon inclusive cross section in the presence of pions
\cite{SST,MTV}.
For the physical, semi-inclusive process, however, its use
would lead to an artificial suppression of the pion-exchange
contribution, especially when the form factor is hard.
The authors of Ref.\cite{SB} also use convolution formulae
within a covariant framework, which, as mentioned above,
inherently makes use of the assumption of factorization as well
as the $k^2$-independence of the off-shell pion structure function,
the justification of which has not yet been demonstrated.

The function ${\cal T}^{S\ s}(t)$ in Eq.(\ref{ope5}) is obtained
by evaluating the trace over the target nucleon spinor and the
Rarita-Schwinger spinor-vector $u_{\alpha}$ for the recoil $\Delta$:
\begin{eqnarray}
{\cal T}^{S\ s}(t)
&=& {\rm Tr} \left[ u(P,S) \bar u(P,S)\
                    u_{\alpha}(p,s) \bar u_{\beta}(p,s)
             \right]
(P - p)^{\alpha} (P - p)^{\beta},
\end{eqnarray}
where \cite{RS}
\begin{mathletters}
\begin{eqnarray}
u_{\alpha}(p,s)
&=& \sum_m \left\langle
           {3 \over 2}\ s \left| 1\ m; {1 \over 2}\ s-m \right.
           \right\rangle
    \epsilon_{\alpha} (m)\ u(p,s-m)
\end{eqnarray}
is constructed from the spin-1/2 Dirac spinor $u$ and spin-1
vectors $\epsilon_{\alpha}(m)$, and normalized such that \cite{BDM}:
\begin{eqnarray}
\sum_s u_{\alpha}(p,s) \bar u_{\beta}(p,s)
&=& \Lambda_{\alpha \beta}(p),
\end{eqnarray}
\begin{eqnarray}
\Lambda_{\alpha \beta}(p)
&=& (\not\!p + M_{\Delta})
\left( - g_{\alpha\beta}
       + {\gamma_{\alpha} \gamma_{\beta} \over 3}
       + {\gamma_{\alpha} p_{\beta} - \gamma_{\beta} p_{\alpha}
         \over 3\ M_{\Delta}}
       + {2\ p_{\alpha} p_{\beta} \over 3\ M_{\Delta}^2}
\right).
\end{eqnarray}
\end{mathletters}%
Because it is emitted collinearly with the pion, production of
$\Delta$ baryons with helicity $\pm 3/2$ is forbidden,
which leads to the selection rule:
\begin{eqnarray}
{\cal T}^{S\ \pm {3 \over 2}}(t) &=& 0.
\end{eqnarray}
This is confirmed by explicit evaluation of the trace
if we recall that for polarized fermion spinors the spin
projection is
$u(P,S) \overline{u}(P,S) = (1 + \gamma_5 \not\!\!\!S)
                            (\not\!\!\!P + M)/2$.
The polarization vectors $S$ and $s$ can be parametrized as:
$S = (0;0,0,+1)$ and
$s = \pm
    \left.
    \left(\sqrt{p_0^2-M_B^2}; p_0 \sin\alpha \cos\phi,
                              p_0 \sin\alpha \sin\phi,
                              p_0 \cos\alpha \right) \right/ M_B$,
so that the angle $\alpha$ between the polarization vectors of the
target proton $S$ and recoiling baryon $s$ coincides with the
direction of the momentum vector ${\bf p}$ relative to the
$z$-axis.
The yield of spin projection $\pm 1/2$\ states is then given by:
\begin{eqnarray}
{\cal T}^{ +{1 \over 2}\ \pm{1 \over 2} }(t)
&=& { 1 \over 12 M_{\Delta}^2 }
    \left[ (M - M_{\Delta})^2 - t \right]\
    \left[ (M + M_{\Delta})^2 - t \right]^2\
    (1 \pm \cos\alpha).
\end{eqnarray}
Because the production of $\Delta$ baryons is limited to
forward angles in the target rest frame,
the factor $(1 \pm \cos\alpha)$
associated with the final state polarization will significantly
suppress the $s = -1/2$ yield relative to that of $s = +1/2$ final
states.

The differential cross section, $Q^2 d^3\sigma / dx dQ^2 d\zeta$,
for the individual polarization states of the produced $\Delta^{++}$
(for DIS from a proton with $S=+1/2$) is shown in Fig.2a for
typical CEBAF kinematics,
$x = 0.14$,  $Q^2 = 2$ GeV$^2$ and $E = 8$ GeV, and in Fig.2b for
$x = 0.075$, $Q^2 = 4$ GeV$^2$ and $E = 30$ GeV, as may be expected
at HERMES.
The pion-exchange model predictions (solid curves)
use the form factor in Eq.(\ref{FF}) with cut-offs
$\Lambda =$ 600 (smallest), 800 and 1000 (largest) MeV,
which gives $< n >_{\pi\Delta} \approx$ 0.01, 0.02 and 0.04,
respectively.
(For $< n >_{\pi\Delta} \approx 0.02$ the cut-off in a $t$-dependent
dipole form factor would be $\sim 700$ MeV.)
The spectrum shows strong correlations between the polarizations
of the target proton ($S=+1/2$) and the $\Delta^{++}$.
In the next Section we examine the extent to which the suppression
of the antiparallel configuration of the $p$ and $\Delta$ spins
in the pion-exchange model is diluted by the competing parton
fragmentation process, which constitutes the main background to the
pion-exchange process discussed here.

\section{BACKGROUNDS}

At the large energy and momentum transfers possible with high-energy
($E \agt 10$ GeV) electron beams, the resonance backgrounds
should not pose a major problem in identifying the required signal.
Firstly, interference from quasi-elastic $\Delta^{++}$ production
will be eliminated by charge conservation.
Secondly, the large $W$ involved means that interference from
excited $\Delta^*$ states (with subsequent decay to $\Delta^{++}$
and pions) will be negligible.
In addition, any such resonance contributions will be strongly
suppressed by electromagnetic form factors at large $Q^2$
($Q^2 \agt 2$ GeV$^2$).

A potentially more significant background will be that due to
uncorrelated spectator fragmentation, as illustrated in Fig.3.
We can estimate the importance of this process within the parton
model framework, in which the cross section is proportional to
(assuming factorization of the $x$ and $\zeta$ dependence
\cite{FF,SSV,SCHM}):
\begin{eqnarray}
\frac{ d^4\sigma^{(s)} }{ dx dQ^2 dz dp_T^2 }
&\propto& {\cal F}_{p\uparrow}(x,Q^2)\
\widetilde{D}_{p\uparrow-q\uparrow\downarrow}^{s}(z,p_T^2),
\end{eqnarray}
where $z = \zeta/(1-x)$ is the light-cone momentum fraction
of the produced baryon carried by the spectator system.
The function ${\cal F}_{p\uparrow}(x,Q^2)$ is proportional to the
spin-weighted interacting-quark momentum distribution functions,
$q^{\uparrow\downarrow} (x) = (q(x) \pm \Delta q(x) )/2$,
where $^{\uparrow \downarrow}$ denote quark spins
parallel or antiparallel to the spin of the proton,
with $q(x)$ and $\Delta q(x)$ being the sum and difference
of $q^{\uparrow}$ and $q^{\downarrow}$, respectively.
For our numerical estimates we use the parametrization of
$\Delta q(x)$ from Gehrmann and Stirling \cite{GS},
and the CTEQ \cite{CTEQ} parametrization for $q(x)$.
The results change little if one uses, for example, the models
of Carlitz and Kaur \cite{CK} or
Sch\"afer \cite{SCHAEFER} for $\Delta q(x)$.

The fragmentation function
$\widetilde{D}_{p\uparrow-q\uparrow\downarrow}^{s}(z,p_T^2)$ gives
the probability for the polarized
($p^{\uparrow}$ minus $q^{\uparrow\downarrow}$)
spectator system to fragment into a $\Delta^{++}$ with
polarization $s$.
The usual assumption is that the transverse momentum distribution
of the baryon also factorizes \cite{SSV,RW,BRE,ACK},\
$\widetilde{D}_{p\uparrow - q\uparrow\downarrow}^s(z,p_T^2)
= D_{p\uparrow - q\uparrow\downarrow}^s(z)\ \varphi(p_T^2)$,
with $\int dp_T^2\ \varphi(p_T^2) = 1$.
To describe the soft, non-perturbative parton fragmentation process,
a number of phenomenological models have been developed for the
fragmentation functions.
Many of these \cite{SLO,BFM} have followed the basic approach
originally formulated by Field and Feynman \cite{FF,FFFRAG},
whose quark jet fragmentation model involved recursive $q \bar{q}$
pair creation (cascade) out of the color field between the scattered
and spectator partons, with subsequent recombination into color
neutral hadrons.
In the original analysis of Ref.\cite{FF} only (unpolarized)
quark $\rightarrow$ meson fragmentation functions were modeled.
Later this approach was extended by Sukhatme et al. \cite{SLO},
and Bartl et al. \cite{BFM} by also allowing for $q \rightarrow$
baryon and $qq \rightarrow$ baryon decays.
The approach pioneered by the Lund group \cite{LUND} included, in
addition, the fragmentation into hadrons of the gluon string
connecting the colored partons.

Analytic expressions for the fragmentation functions can be obtained
by constraining their limiting behavior at the asymptotic limits.
The $z \rightarrow 0$ limit requires a $1/z$ behavior for $D(z)$ in
order to reproduce the observed logarithmic increase in hadron
multiplicity as $s \rightarrow \infty$,
\begin{eqnarray}
<N_B> &=& \int_{z_{min}}^1\ dz\ D(z)\ \ \sim\ \ \ln s,
\label{nB}
\end{eqnarray}
where $z_{min} \propto 1/s$ (see below).
For the $z \rightarrow 1$ limit one commonly applies dimensional
counting rules \cite{FPS}, using essentially the same arguments
as for the $x \rightarrow 1$ limit of structure functions \cite{COUNT}.
For the specific case of the $\Delta^{++}$, at large $z$ this should
carry most of the parent system's momentum, and therefore contain
both valence $u$ quarks from the target proton.
In our region of interest, namely $z \agt 0.6$, where the pionic
contribution is the largest, by far the most important contributions
to $D(z)$ come from the process whereby the $\Delta^{++}$ is formed
after only one $u \bar{u}$ pair is created \cite{SLO,BFM}.
As a consequence, DIS from valence $u$ quarks will not be too
important.
For scattering from sea quarks we assume the same fragmentation
probabilities for $uuq\bar q$ spectator states as for $uu$,
although in general multi-quark configurations could decay at
different rates than the valence diquark
(however, already at $Q^2 \simeq 2$ GeV$^2$ the sea constitutes
at most $\sim 15\%$ of the cross section at $x \sim 0.1$).

Rather than rely on model counting rule arguments,
we parametrize the (very limited) EMC data \cite{EMC86}
on unpolarized $\Delta^{++}$ muon production
for $z \rightarrow 1$ as:
$D_{uu}(z \rightarrow 1) = \alpha (1 - z)^{\beta}$,
where $\beta \approx 0.3$.
The overall normalization of the fragmentation function
is fixed by the data to be
$\alpha \approx 0.68$.
Note that in obtaining this parametrization it has been necessary
to perform a conversion of the kinematic variables.
Usually in semi-inclusive experiments \cite{EMC86} the longitudinal
momentum dependence is measured as a function of the Feynman variable
$x_F$, defined as the ratio of the center of mass longitudinal
momentum to its maximum allowed
value, $x_F = p^*_L / p^*_{L max}
       \simeq 2 |{\bf p}^*| / \sqrt{W^2}
       \simeq 1 - M_X^2 / W^2$,
where $M_X$ is the mass of the inclusive hadronic debris,
and the asterisk ($^*$) denotes center of mass momenta.
This variable can be related to the light-cone variable $z$ via
\begin{eqnarray}
z &=& { \sqrt{ M_{\Delta}^2 + p_T^2 + W^2\ x_F^2/4 }
      - \sqrt{W^2}\ x_F/2
        \over \sqrt{W^2} }.
\end{eqnarray}
Note that for $z \rightarrow 1$, $x_F \simeq z$
if $W^2 \gg M_{\Delta}^2 + p_T^2$.
The target (current) fragmentation region corresponds to
$x_F < 0$ ($x_F > 0$), and the boundary between the regions
at $x_F = 0$ corresponds to
$\zeta_{min} = \sqrt{M_{\Delta}^2 + p_T^2} / \sqrt{W^2}$.

To model the spin dependence of the fragmentation process we follow
the simple approach taken by Bartl et al. \cite{BFMS} (see also
Refs.\cite{BIGI,DON}) in their study of polarized quark $\rightarrow$
baryon fragmentation.
Namely, the diquark is assumed to retain its helicity
during its decay, and the $q \bar{q}$ pair creation probability
is independent of the helicity state of the quark $q$.
At leading order this means that the produced baryon contains
the helicity of the diquark, so that, for example,
a $\Delta^{\Uparrow}$ or $\Delta^{\uparrow}$ can
emerge from a $q^{\uparrow} q^{\uparrow}$ diquark,
whereas a $\Delta^{\Downarrow}$ cannot.
(Our notation here is that
$\Uparrow, \uparrow, \downarrow, \Downarrow$
represent $s = +3/2, +1/2, -1/2, -3/2$ states, respectively.)

The overall normalization of the spin-dependent fragmentation
functions is fixed by the condition
\begin{eqnarray}
         q(x)\ D_{p-q}(z)\
+\ \bar{q}(x)\ D_{p-\bar{q}}(z)\
&=& q^{\uparrow} (x)\
    D_{p\uparrow - q\uparrow}(z)\
 +\ q^{\downarrow} (x)\
    D_{p\uparrow - q\downarrow}(z)\       \nonumber\\
&+& \bar{q}^{\uparrow} (x)\
    D_{p\uparrow - \bar{q}\uparrow}(z)\
 +\ \bar{q}^{\downarrow} (x)\
    D_{p\uparrow - \bar{q}\downarrow}(z),
\label{norm}
\end{eqnarray}
where
\begin{eqnarray}
D(z)
&=& \sum_{s=-3/2}^{+3/2} D^{s}(z).
\end{eqnarray}
In relating the production rates for various polarized $\Delta^{++}$
we employ SU(6) spin-flavor wave functions, from which
simple relations among the valence diquark $\rightarrow$
$\Delta^{++}$ fragmentation functions, $D_{ qq_{j(j_z)} }^{s}(z)$,
can be deduced (the diquark state $qq_{j(j_{z})}$ is labeled by
its spin $j$ and spin projection $j_z$).
The leading functions are related by:
\begin{eqnarray}
D_{ uu_{1(1)} }^{\Uparrow}(z)
= 3\ D_{ uu_{1(1)} }^{\uparrow}(z)
= \frac{3}{2} D_{ uu_{1(0)} }^{\uparrow}(z)
= \frac{3}{2} D_{ uu_{1(0)} }^{\downarrow}(z),  \label{leadFF}
\end{eqnarray}
with normalization determined from:
\begin{eqnarray}
D_{ uu_{1(1)} }^{\Uparrow}(z)
&=& \frac{3}{4} D_{ uu }(z).                    \label{polunp}
\end{eqnarray}
(Note that this is true only when the spin projections of the diquark
and $\Delta$ are aligned.)
The non-leading fragmentation functions are those which require
at least two $q \bar{q}$ pairs to be created from the vacuum, namely
$D_{ uu_{1(0)} }^{\Uparrow/\Downarrow}$,
$D_{ uu_{1(1)} }^{\downarrow/\Downarrow}$,
$D_{ ud_{0(0)} }^{\Uparrow/\uparrow/\downarrow/\Downarrow}$,
$D_{ ud_{1(0)} }^{\Uparrow/\uparrow/\downarrow/\Downarrow}$,
and
$D_{ ud_{1(1)} }^{\Uparrow/\uparrow/\downarrow}$,
and those which require 3 such pairs,
$D_{ uu_{1(1)} }^{\Downarrow}$ and
$D_{ ud_{1(1)} }^{\Downarrow}$.
Except at very small $z$ ($z \alt 0.2$) the latter functions
are consistent with zero \cite{BFM}.
For the 2-$q \bar{q}$ pair fragmentation functions, we also expect that
$ D_{ uu_{1(0)} }^{\Uparrow}(z)
= D_{ uu_{1(0)} }^{\Downarrow}(z)$.
For $z \agt 0.2$ the unpolarized model fragmentation functions
of Ref.\cite{BFM} requiring two $q \bar{q}$ pairs (e.g. $D_{ud}(z)$)
are quite small compared with the leading fragmentation functions,
$D_{ud}(z) \simeq 0.1\ D_{uu}(z)$.
For spin-dependent fragmentation we therefore expect
a similar behavior for those decay probabilities requiring two
$q \bar{q}$ pairs created in order to form the final
state with the correct spin and flavor quantum numbers.
This then allows for a complete model description of the
polarized fragmentation backgrounds
at large $z$ in terms of only the 4 fragmentation functions in
Eq.(\ref{leadFF}).

Finally, the $p_T$-integrated differential cross section
for the electroproduction of a $\Delta^{++}$ with spin $s$
can be written:
\begin{eqnarray}
{ d^3\sigma^{(s)} \over dx dQ^2 d\zeta }
&=&
\left( { 2 \pi \alpha^2  \over  M^2 E^2 x (1-x) } \right)
\left( { 1 \over 2 x^2 }\
    +\ { 4 M^2 E^2 \over Q^4 }
       \left( 1 - {Q^2 \over 2 M E x} - { Q^2 \over 4 E^2 } \right)
\right)
\label{qpmful}\\
& & \hspace*{-1.3cm} \times
\left[
{4 x \over 9}
\left( u_V^{\uparrow} D_{ud_{1(0)}}^{s}
     + 2 \bar{u}^{\uparrow} \left({2 \over 3} D_{uu_{1(1)}}^{s}
                           + {1 \over 3} D_{uu_{1(0)}}^{s}
                       \right)
     + u_V^{\downarrow} D_{ud_{1(1)}}^{s}
     + 2 \bar{u}^{\downarrow} \left({2 \over 3} D_{uu_{1(1)}}^{s}
                           + {1 \over 3} D_{uu_{1(0)}}^{s}
                       \right)
\right)
\right.
\nonumber\\
& & \hspace*{-1.0cm} +
\left.
{x \over 9}
\left( d_V^{\uparrow} D_{uu_{1(0)}}^{s}
     + 2 \bar{d}^{\uparrow} \left({2 \over 3} D_{uu_{1(1)}}^{s}
                           + {1 \over 3} D_{uu_{1(0)}}^{s}
                       \right)
     + d_V^{\downarrow} D_{uu_{1(1)}}^{s}
     + 2 \bar{d}^{\downarrow} \left({2 \over 3} D_{uu_{1(1)}}^{s}
                           + {1 \over 3} D_{uu_{1(0)}}^{s}
                       \right)
\right)
\right].
\nonumber
\end{eqnarray}

In Figs.2a and 2b the parton model predictions (dashed) for the various
polarization states of the $\Delta$ are plotted in comparison
with the pion-exchange cross sections.
In the quark-parton model the correlations are significantly weaker,
with the ratio of polarized $\Delta$s being
$s = +3/2 : +1/2 : -1/2 : -3/2 \approx 3 : 2 : 1 : 0$.
The comparisons assume that there is no significant interference
between the parton fragmentation and pion-exchange contributions.
At small values of the exchanged four-momentum squared $t$ one may
expect this to be a good approximation, since the distance scales
at which the pion and diquark are formed are rather different.
For larger values of $t$ this approximation may be less justifiable,
and the possibility would exist that interference effects could
modify the above simple predictions.
This problem would be most pronounced for hard $\pi N \Delta$
vertices, however for relatively soft form factors
($\Lambda \alt 700$ MeV) the above predictions should be a reliable
guide.

Although the total unpolarized parton model cross sections
are larger than the pion-exchange cross sections, even at
larger values of $\zeta$ where the pionic effects are
strongest, the polarization aligned component in the pion
model is larger than that in the parton model.
The differences between the pion-exchange model and fragmentation
backgrounds can be further enhanced by examining polarization
asymmetries.
In Fig.4 we show the difference $\sigma^+ - \sigma^-$,
where
$\sigma^{\pm} \equiv Q^2 d^3\sigma^{(s=\pm 1/2)} / dx dQ^2 d\zeta $,
as a fraction of the total unpolarized cross section,
for the two kinematic cases in Figs.2a and 2b
(solid = CEBAF kinematics;
dashed = HERMES kinematics).
The resulting $\zeta$ distributions are almost flat, but
significantly different for the two models ($\pi$ and $qq$ label
the pion-exchange and spectator diquark fragmentation models).
We have also calculated the ratio for the form factor in
Eq.(\ref{FFt}), and find the results to be almost
indistinguishable from those in Fig.4.
Therefore, a measurement of the polarization asymmetry appears to
test only the presence of a pionic component of the nucleon
wave function, independent of the details of the form factor.

Of course the two curves in Fig.4 represent extreme cases,
in which $\Delta$s are produced entirely via pion emission or
diquark fragmentation.
In reality we can expect a ratio of polarization cross sections
which is some average of the curves in Fig.4.
The amount of deviation from the parton model curve will indicate
the extent to which the pion-exchange process contributes.  From
this, one can in turn deduce the strength of the $\pi N \Delta$
form factor.
Unlike inclusive DIS, which can only be used to place upper bounds
on the pion number, the semi-inclusive measurements could pin down
the absolute value of $< n >_{\pi\Delta}$.
A measurement of this ratio would thus be particularly useful
in testing the relevance of non-perturbative degrees of freedom
in high energy processes.

\section{Kaon Cloud of the Nucleon}

Semi-inclusive leptoproduction of polarized $\Lambda$ hyperons from
polarized protons can also be used to test the relevance
of a kaon cloud in the nucleon, Fig.5.
The advantage of detecting $\Lambda$s in the final state, as compared
with $\Delta$ baryons lies in the fact that the $\Lambda$ is
self-analyzing.
It has, in fact, been suggested recently (see Ref.\cite{HMC}) that
measurement of the polarization of the $\Lambda$ in the target
fragmentation region could discriminate between models of the spin
content of the nucleon, in which a large fraction of the spin is
carried either by (negatively polarized) strange quarks or
(positively polarized) gluons.
The latter would imply a positive correlation of the
target proton and $\Lambda$ spins, while the spin projection
of the $\Lambda$ along the target polarization axis should be
negative in the former model.
(Similar effects would also be seen in the reaction
$\bar p p \rightarrow \bar \Lambda \Lambda$ \cite{AEK}.)
A kaon cloud would be the natural way to obtain a polarized strange
sea of the proton.

Although some data do exist for $\Lambda$ production in the region
$x_F < 0$ \cite{LAMDAT}, the large errors and limited range of $x_F$
do not permit one to unambiguously discern the presence of $K$ effects.
A direct test of the presence of a kaon cloud of the nucleon would be
to observe the differential $\Lambda$ production cross section
at large $\zeta$, and in particular the relative polarization yields.

The formalism for the DIS off the kaon cloud \cite{TTT} is very similar
to that for the pion exchange model in Section III.
Included in the observed $\Lambda$ cross section will be
contributions from direct $\Lambda$ production via $K^+$ exchange,
as well as those from $\Sigma^0$ recoil states, which subsequently
decay to $\Lambda \gamma$.
The differential hyperon $H$ ($= \Lambda, \Sigma^0$)
production cross section is similar to that in Eq.(\ref{ope5}), with
the trace factor here given by:
\begin{eqnarray}
\label{traceL}
{\cal T}_{H}^{S\ s}(t)
&=& \left( P \cdot p\ -\ M M_{H} \right)
    \left( 1 + S \cdot s \right)\
 -\ P \cdot s\ p \cdot S.
\end{eqnarray}
Using the rest frame parametrizations of the individual momentum
and spin vectors, we find for a proton initially polarized in the
positive $z$ direction:
\begin{eqnarray}
{\cal T}_{H}^{+{1\over 2}\ \pm{1\over 2}}(t)
&=& { 1 \over 2 } \left( (M_{H}-M)^2 - t \right)
    \left( 1 \pm \cos\alpha \right).
\end{eqnarray}

The $\Lambda$ production cross section is shown in Fig.6,
as a function of $\zeta$, for the two possible polarizations
(the kinematics are as in Fig.2a).
The $K$-exchange predictions are calculated for the form factor
in Eq.(\ref{FF}) with cut-offs of $\Lambda = 0.6$ (smallest),
0.8 and 1.0 GeV (largest).
The $K$-exchange model predicts very strong correlations
between the target and recoil polarizations, so that the
asymmetry shown in Fig.7 for the cross sections
$(\sigma^+ - \sigma^-) / (\sigma^+ + \sigma^-)$,
where $\sigma^{\pm} \equiv Q^2 d^3\sigma^{(s=\pm 1/2)}/dx dQ^2 d\zeta$,
is almost unity.
The $K$-exchange ratios are very similar to the $\pi$-exchange
results in Fig.4, indicating the similar spin transfer dynamics
inherent in the meson cloud picture of the nucleon.

This is in strong contrast with the expectation from the
$qq \rightarrow \Lambda$ diquark fragmentation picture,
in which the target--recoil spin correlation is much weaker.
In fact, to a first approximation the $\Lambda^{\uparrow\downarrow}$
yields in the quark-parton model are equal.
Assuming an SU(6) symmetric wave function for the $\Lambda$,
namely $\Lambda^{\uparrow\downarrow}
\sim s^{\uparrow\downarrow} (ud)_{spin=0}$,
the leading fragmentation function will be
$D_{ud_{0(0)}}^{\Lambda^{\uparrow\downarrow}}$,
so that the relevant component of the SU(6) proton wave function
is $u^{\uparrow} (ud)_{spin=0}$.
Since one has equal probabilities to form a $\Lambda^{\uparrow}$
and $\Lambda^{\downarrow}$, in the leading fragmentation
approximation the asymmetry will be zero.
Of course, SU(6) symmetry breaking effects, as well as non-leading
fragmentation contributions, will modify this result, as will
contributions from the production and decay of
$\Sigma^{0 \uparrow\downarrow}$ hyperons (from the SU(6)
wave function one can see that a $\Sigma^{0 \uparrow}$ is more
likely to form from a $p^{\uparrow}$ than is a
$\Sigma^{0 \downarrow}$).
However, the qualitative result that the asymmetry is small should
remain true.
Therefore the observation of a large polarization asymmetry in the
large-$\zeta$ region of the target fragmentation region will be
evidence for a kaon-exchange fragmentation mechanism.

\section{OTHER TESTS OF PION EXCHANGE}

In addition to the above described process which may be tested
in upcoming experiments, several other novel ideas have been
proposed to identify a pionic component of the nucleon wave function.
In this section we will briefly outline a couple of them.

\subsection{Exclusive electroproduction of pions}

In a recent detailed analysis \cite{HEID}, it has been
suggested that measurements of fast pions in the final state
in coincidence with the final electron could be sensitive to
a pionic component of the nucleon.
Extending the exclusive analysis of G\"uttner et al. \cite{GCPP} in
the IMF, Pirner and Povh \cite{HEID} work within a constituent
quark picture in which the probability to find a pion in the
nucleon is expressed in terms of the pion distribution function
inside a constituent quark.

The differential pion-production cross section for the
``leading pion'' (integrated over transverse momenta)
is written as:
\begin{eqnarray}
{d\sigma \over dx dy dz}
&\propto& A(x,y,z)\ +\ B(x,y,z) F_{\pi}(Q^2)\
                    +\ C(x,y,z) F_{\pi}^2(Q^2),
\label{piprod}
\end{eqnarray}
where $z = E_{\pi}/\nu$ is the fraction of the photon's energy
carried by the pion,
and where the $A$ and $B$ terms describe soft and hard fragmentation,
respectively.
The function $C$ reflects coherent scattering from the
pion cloud of the constituent quark.
Each term in Eq.(\ref{piprod}) gives a characteristic $Q^2$-dependence,
namely $\log Q^2$, $1/Q^2$ and $1/Q^4$, respectively.
To isolate the coherent scattering from the pion one therefore
has to restrict oneself to the a region of not too high $Q^2$,
where the form factor suppression has not yet eliminated the
pion signal.

The useful observation in this analysis is that each of the
three processes has a quite distinct $z$-dependence.
The hard-fragmentation process gives a differential
cross section which is constant in $z$, and is important
in the intermediate $z$ region ($0.6 \alt z \alt 0.8$).
The soft fragmentation mechanism is dominant at small $z$,
$z \alt 0.6$, but dies out rather rapidly at larger $z$.
This fact may enable one to detect the pion-exchange process,
which dominates the region $0.8 \alt z \alt 1$, where it
predicts a contribution that is several times larger than
the constant-$z$, hard fragmentation mechanism.
The conclusion that the pion-exchange process is dominant is
consistent with our results above.

\subsection{Semi-inclusive scattering from $^3He$.}

Another novel idea was recently put forward by Dieperink and Pollock
\cite{DP}, where the suggestion was to measure the recoiling $^3He$
nucleus in deep-inelastic scattering from a $^3He$ target at
$x \sim 0.05 - 0.1$.
Unlike the exclusive experiments, one would not need to restrict
oneself to the small $Q^2$ region.
Because of the rather small probability for the nucleus to remain
intact after a hard interaction with a parton in one of the
constituent nucleons, backgrounds due to parton fragmentation
would be virtually eliminated.
Therefore the most likely mechanism responsible for the final state
$^3He$ nucleus would be DIS from a non-nucleonic component in the
target, the typical candidate being a pion.
Other non-nucleonic constituents could also give rise to
the same final state, such as DIS from a Pomeron in the diffractive
region.
However, in practice these could be eliminated by restricting oneself
to the specific kinematic region of not too small $x$.

An additional problem here in obtaining unambiguous information about
the $N \pi$ vertex from the $^3He\ \pi$ vertex would be nuclear effects
in the $^3He$ nucleus.
Furthermore, any final state interactions, leading to the
break up of the $^3He$ nucleus, could decrease the apparent number
of pions seen in the reaction, thus leading to an underestimate of
the pion multiplicity in the nucleon.
Nevertheless, this is an interesting idea, and a detailed study
should be performed with a view to determining the feasibility of
conducting this experiment in future.

\section{CONCLUSION}

We have outlined a series of semi-inclusive experiments on polarized
proton targets that may for the first time enable one to unambiguously
establish the presence of a pion and kaon cloud of the nucleon at
high energies.
The most difficult part of the calculation is the estimate of the
size of the competing diquark fragmentation process.
While we have used as much experimental data and theoretical
guidance as possible, in order to make that calculation reliable,
it could undoubtedly benefit from further study into the
polarized diquark $\rightarrow$ polarized baryon fragmentation
process.
Even bearing this caution in mind, our results (especially
Figs.4 and 7) are extremely encouraging.
While the experiments proposed here are difficult, requiring
all the intensity and duty factor one can obtain with modern
electron accelerators, it does seem that they will provide quite
clear information on the role of the pseudoscalar mesons in the
nucleon.

\acknowledgements

We would like to thank W. Weise for a careful reading of the manuscript.
This work was partially supported by the
Australian Research Council
and the BMFT.
W.M. would like to thank the University of Adelaide for its
hospitality during a recent visit, where this work was completed.


\newpage

\begin{figure}
\centering{\ \psfig{figure=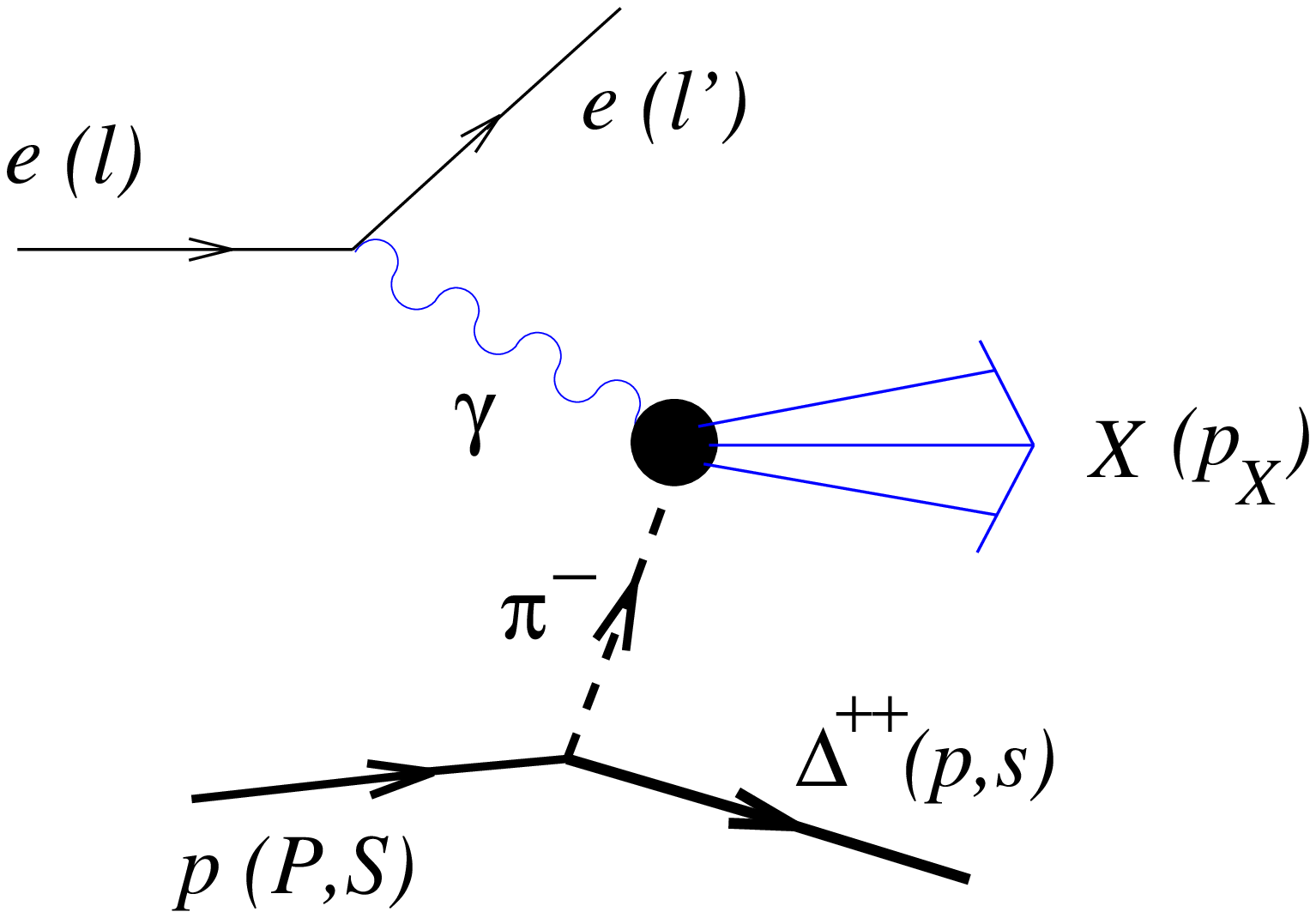,height=11cm}}
\caption{Pion-exchange model of the semi-inclusive deep-inelastic
         scattering from a polarized proton with a polarized recoil
         $\Delta^{++}$ in the final state.}
\label{F1}
\end{figure}

\begin{figure}
\centering{\ \psfig{figure=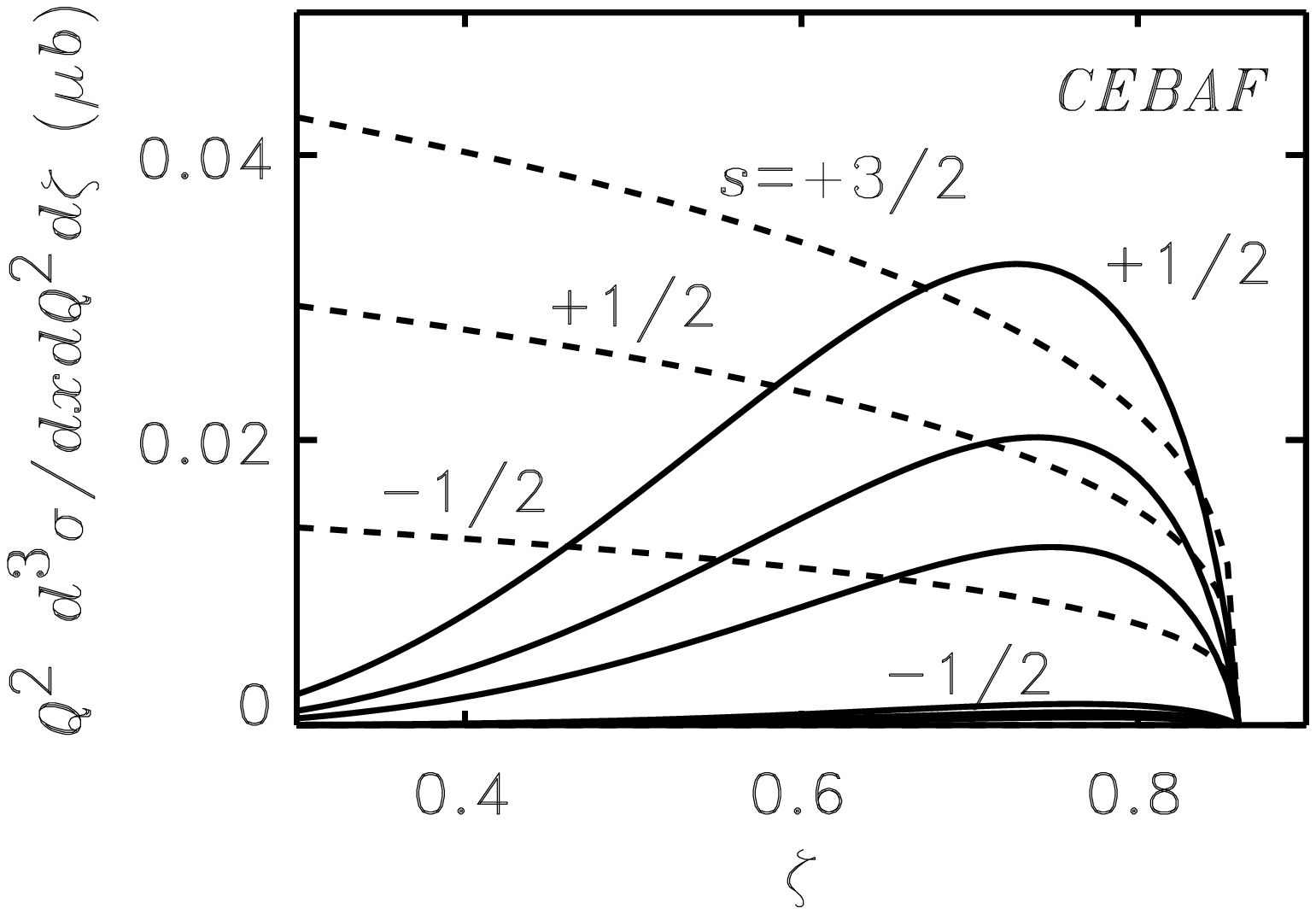,height=8cm}}
\centering{\ \psfig{figure=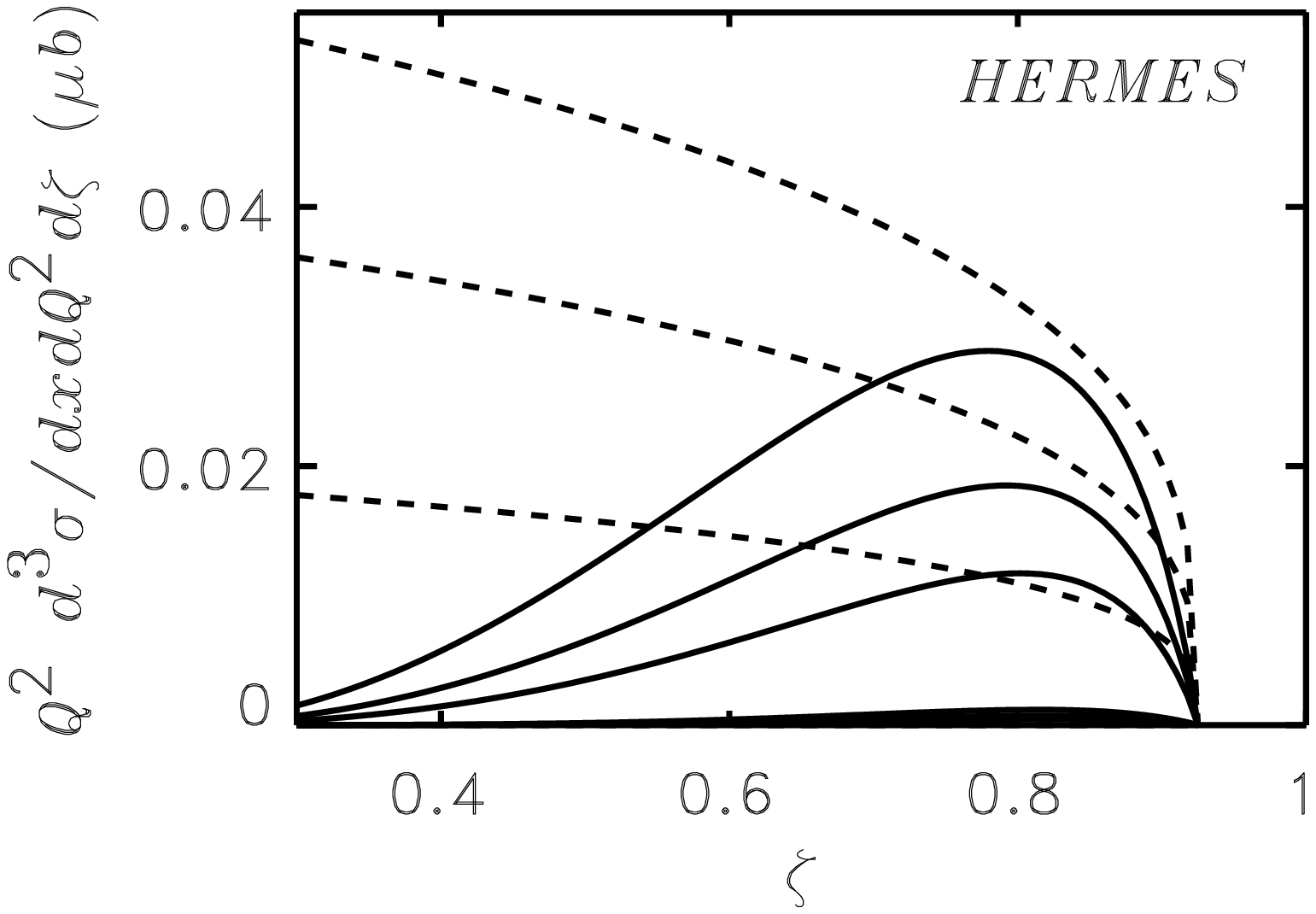,height=8cm}}
\caption{Differential electroproduction cross section for various
         polarization states of the $\Delta^{++}$, for typical
         (a) CEBAF and (b) HERMES kinematics (see text).
         The $\pi$-exchange model predictions (solid) are for cut-off
         masses $\Lambda =$ 600 (smallest), 800 and 1000 (largest) MeV.
         The top three solid curves are for spin $s=+1/2$ final states,
         while the bottom three solid curves are for $s=-1/2$.
         The quark-parton model background (dashed) is estimated using
         the fragmentation functions extracted from the unpolarized
         EMC data \protect\cite{EMC86} and Eqs.(\protect\ref{leadFF})
         and (\protect\ref{polunp}).}
\label{F2}
\end{figure}

\begin{figure}
\centering{\ \psfig{figure=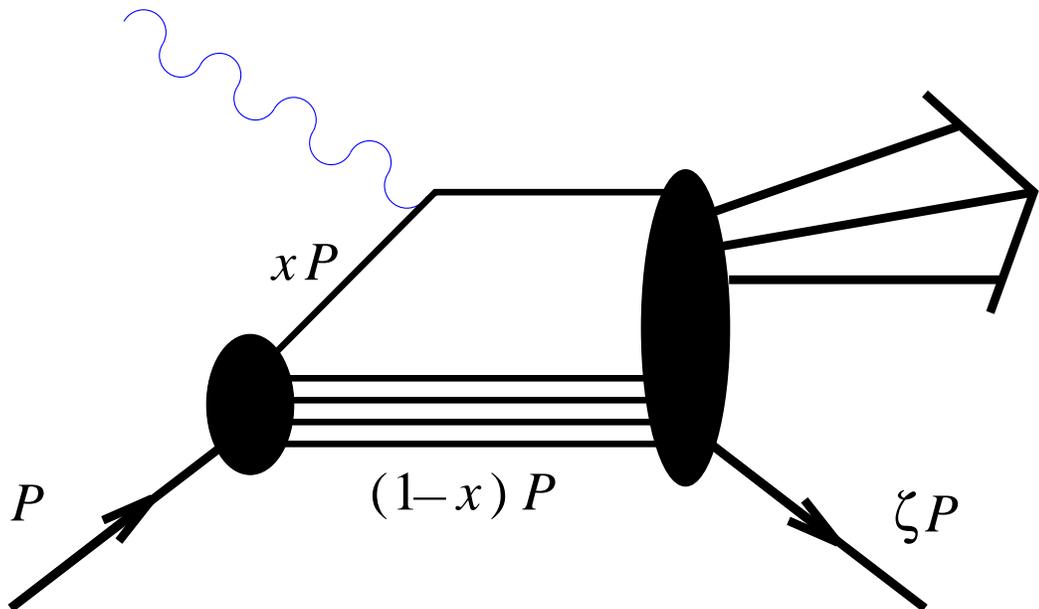,height=10cm}}
\caption{Background parton (spectator `diquark') fragmentation
         process leading to the same $\Delta^{++}$ final state.}
\label{F3}
\end{figure}

\newpage
\begin{figure}
\centering{\ \psfig{figure=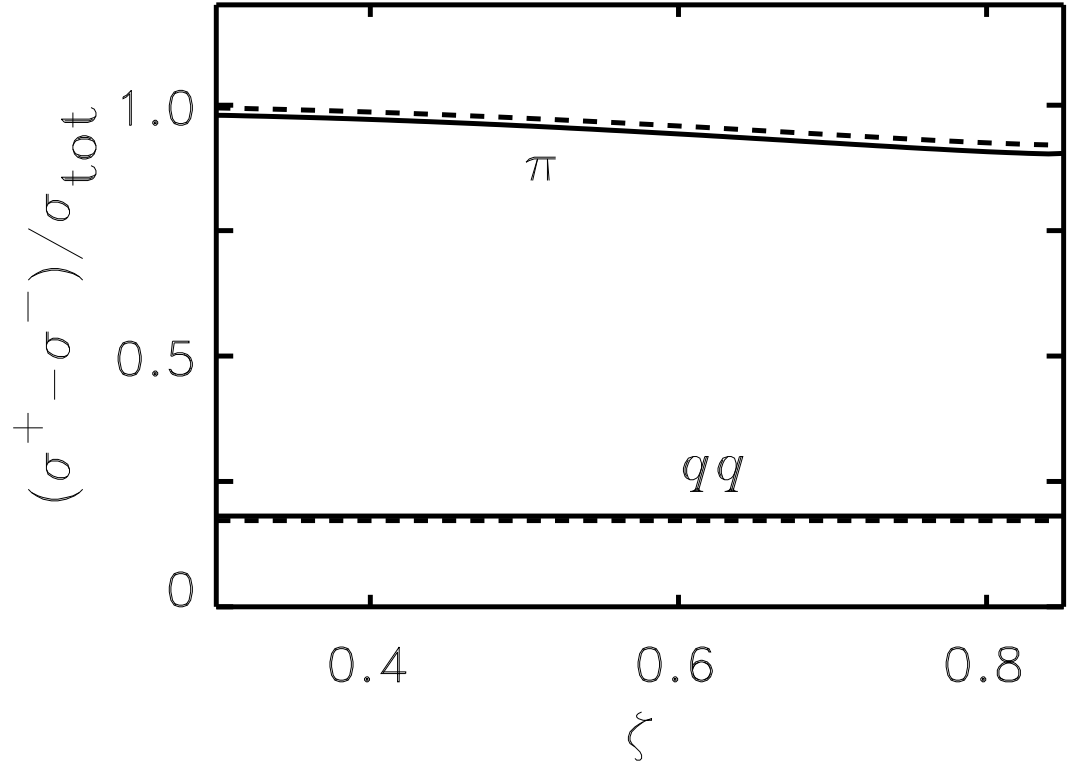,height=12cm}}
\caption{Polarization asymmetry for the $\pi$-exchange (upper curves)
         and parton fragmentation (lower curves) models,
         with $\sigma^{\pm}$ as defined in the text, and
         $\sigma_{\rm tot}$ is the sum over all polarization states.
         The solid and dashed lines are for CEBAF and HERMES
         kinematics, respectively.}
\label{F4}
\end{figure}

\newpage
\begin{figure}
\centering{\ \psfig{figure=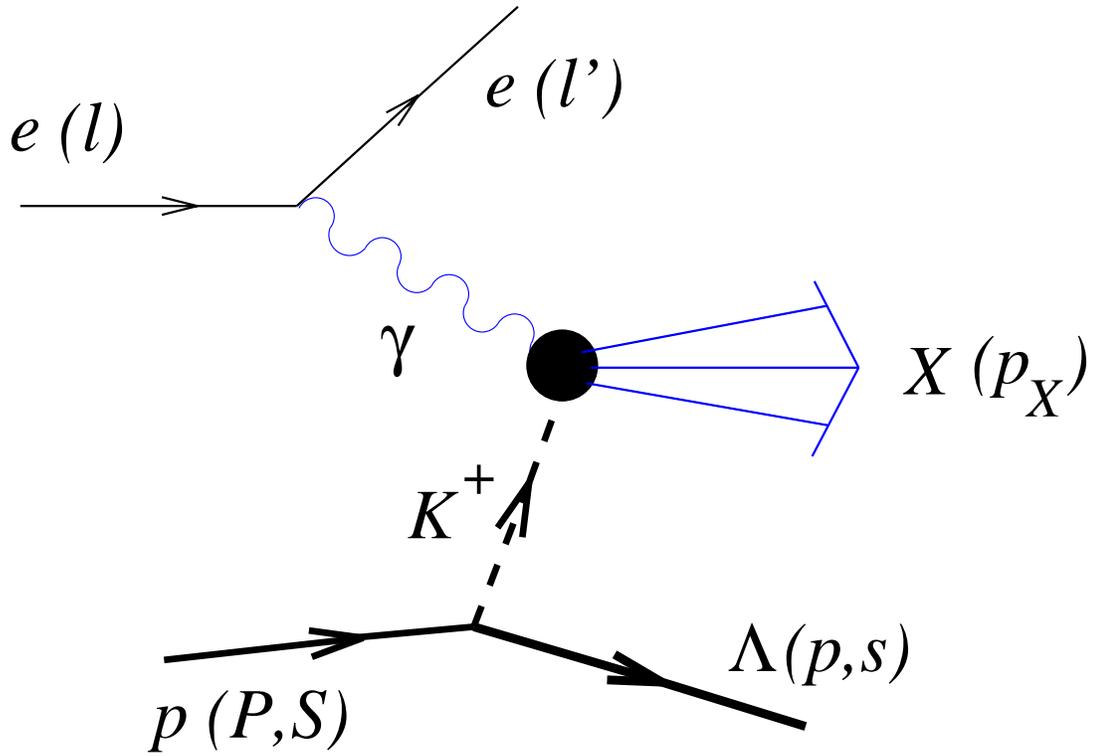,height=11cm}}
\caption{Kaon-exchange mechanism for the semi-inclusive production of
         polarized $\Lambda$ hyperons.}
\label{F5}
\end{figure}

\newpage
\begin{figure}
\centering{\ \psfig{figure=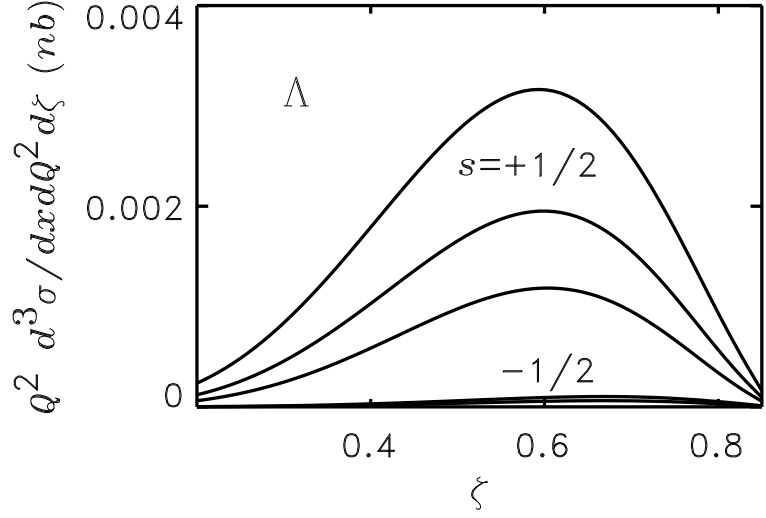,height=8cm}}
\caption{Differential $\Lambda$ production cross section in the
         $K$-exchange model, for form factor cut-offs
         $\Lambda =$ 600 (smallest), 800 and 1000 (largest) MeV,
         for $s=+1/2$ (upper three curves) and $s=-1/2$ (lower
         three curves) final states.
         Also included are contributions from $K^+ \Sigma^0$
         states, with the subsequent decay
         $\Sigma^0 \rightarrow \Lambda \gamma$.}
\label{F6}
\end{figure}

\begin{figure}
\centering{\ \psfig{figure=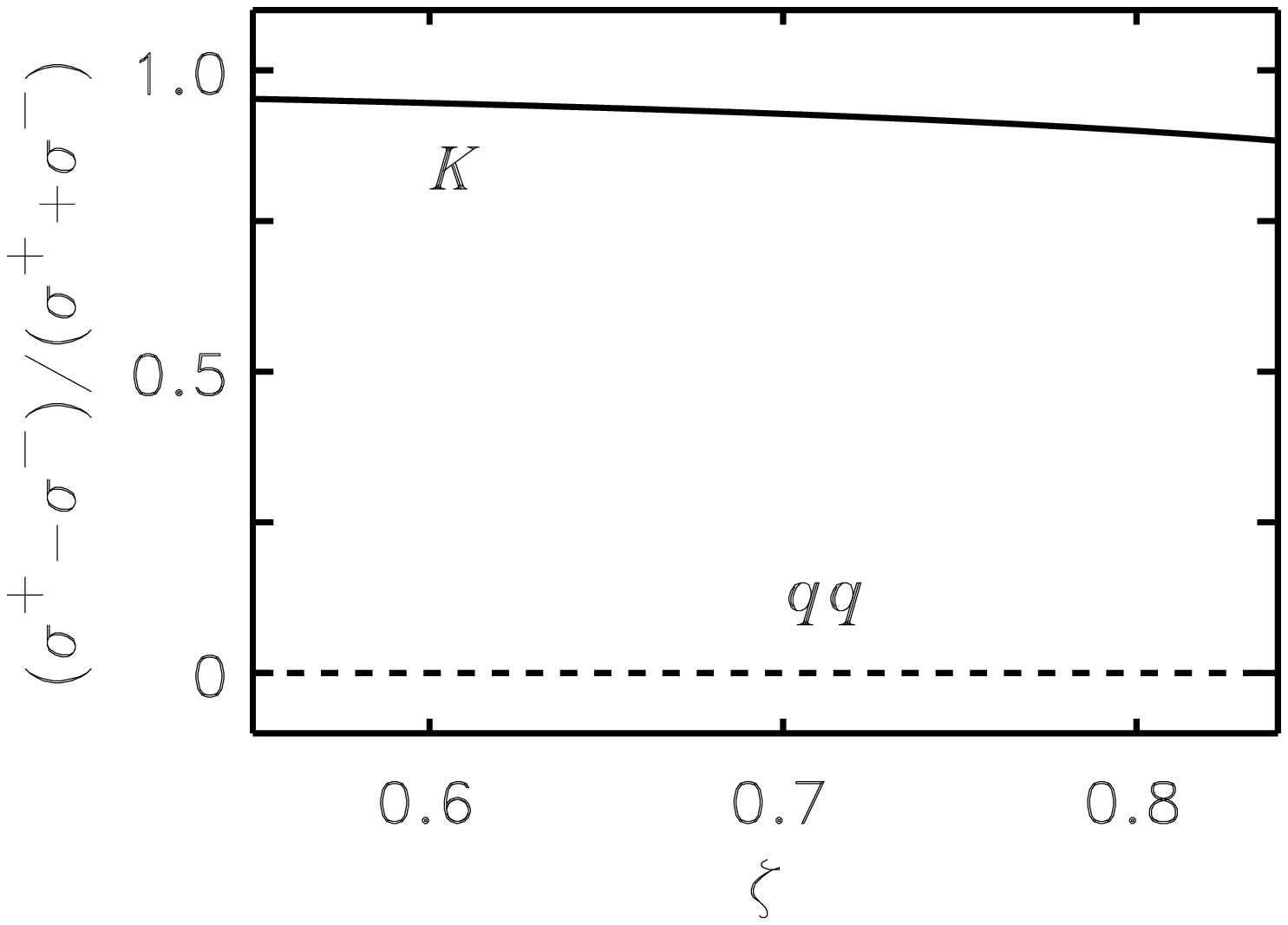,height=8cm}}
\caption{Polarization asymmetry for the $K$-exchange (solid)
         model of $\Lambda$ production, compared with a leading
         fragmentation approximation estimate for the parton
         fragmentation process (dashed).}
\label{F7}
\end{figure}


\references

\bibitem{GSR}   Gottfried, K.:
                Phys.Rev.Lett. {\bf 18}, 1174 (1967).

\bibitem{NMC}   Amaudruz, P., et al. (New Muon Collaboration):
                Phys.Rev.Lett. {\bf 66}, 2712 (1991);
                Phys.Rev. D {\bf 50}, 1 (1994).

\bibitem{HM}    Henley, E.M. and Miller, G.A.:
                Phys.Lett. B {\bf 251}, 497 (1990).

\bibitem{SST}   Signal, A.I., Schreiber, A.W. and Thomas, A.W.:
                Mod.Phys.Lett. A {\bf 6}, 271 (1991).

\bibitem{MTS}   Melnitchouk, W., Thomas, A.W. and Signal, A.I.:
                Z.Phys. A {\bf 340}, 85 (1991).

\bibitem{KL}    Kumano, S. and Londergan, J.T.:
                Phys.Rev. D {\bf 44}, 717 (1991).

\bibitem{MA}    Ma, B.-Q., Sch\"afer, A. and Greiner, W.:
                Phys.Rev. D {\bf 47}, 51 (1993).

\bibitem{JUL}   Hwang, W.-Y.P., Speth, J. and Brown, G.E.:
                Z.Phys. A {\bf 339}, 383 (1991);
%
                Szczurek, A. and Speth, J.:
                Nucl.Phys. {\bf A555}, 249 (1993).

\bibitem{NA51}  Baldit, A., et al. (NA51 Collaboration):
                Phys.Lett. B {\bf 332}, 244 (1994).

\bibitem{EMCG1} Ashman, J., et al. (European Muon Collaboration):
                Nucl.Phys. {\bf B328}, 1 (1989).

\bibitem{SMC}   Adeva, B., et al. (Spin Muon Collaboration):
                Phys. Lett. B {\bf 329}, 399 (1994).

\bibitem{SLACG1} Abe, K., et al. (E143 Collaboration):
                Phys.Rev.Lett. {\bf 74}, 346 (1995).

\bibitem{U1GLU} Altarelli, G. and Ross, G.G.:
                Phys.Lett. B {\bf 212}, 391 (1988);
%
                Carlitz, R.D., Collins, J.C. and Mueller, A.H.:
                Phys.Lett. B {\bf 214}, 229 (1988);
%
                Efremov, A.V. and Teryaev, O.V.:
                Dubna preprint JINR-E2-88-287 (1988);
%
                Jaffe, R.L. and Manohar, A.:
                Nucl.Phys. {\bf B337}, 509 (1990);
%
                Steininger, K. and Weise, W.:
                Phys.Rev. D {\bf 48}, 1433 (1993);
%
                Bass, S.D. and Thomas, A.W.:
                Prog.Part.Nucl.Phys. {\bf 33}, 449 (1994);
%
                Narison, S., Shore, G.M. and Veneziano, G.:
                Nucl.Phys. {\bf B433}, 209 (1995).

\bibitem{TTT}   Signal, A.I. and Thomas, A.W.:
                Phys.Lett. B {\bf 191}, 205 (1987).

\bibitem{EJ}    Ellis, J. and Jaffe, R.L.:
                Phys.Rev. D {\bf 9}, 1444 (1974).

\bibitem{SUL}   Sullivan, J.D.:
                Phys.Rev. D {\bf 5}, 1732 (1972).

\bibitem{T83}   Thomas, A.W.:
                Phys.Lett. {\bf 126} B, 97 (1983).

\bibitem{FMS}   Frankfurt, L.L., Mankiewicz, L. and Strikman, M.I.:
                Z.Phys. A {\bf 334}, 343 (1989).

\bibitem{HEID}  Pirner, H.J. and Povh, B.:
                in Proceedings of the Italian Physical Society,
                Vol.44 (The ELFE project: an electron laboratory
                for Europe),
                ed. J. Arvieux and E. De Santis, 1992.

\bibitem{DP}    Dieperink, A.E.L. and Pollock, S.J.:
                Z.Phys. A {\bf 348}, 117 (1994).

\bibitem{STA}   Stack, J.D.:
                Phys.Rev.Lett. {\bf 28}, 57 (1972).

\bibitem{LUS}   Lusignoli, M. and Srivastava, Y.:
                Nucl.Phys. {\bf B138}, 151 (1978);
%
                Lusignoli, M., Pistilli, P., and Rapuano, F.:
                Nucl.Phys. {\bf B155}, 394 (1979).

\bibitem{CEBAF} Melnitchouk, W. and Thomas, A.W.:
                in Proceedings of the Workshop on CEBAF at
                Higher Energies,
                eds. N.Isgur and P.Stoler (April 1994) p.359.

\bibitem{BEB77} Bebek, C.J., et al.:
                Phys.Rev. D {\bf 15}, 3077 (1977).

\bibitem{ARN85} Arneodo, M., et al. (EM Collaboration):
                Phys.Lett. B {\bf 150}, 458 (1985).

\bibitem{EMC86} Arneodo, M., et al. (EM Collaboration):
                Nucl.Phys. {\bf B264}, 739 (1986).

\bibitem{E745}  Kitagaki, T., et al. (E745 Collaboration):
                Phys.Lett. B {\bf 214}, 281 (1988).

\bibitem{BEBC}  Guy, J., et al. (BEBC Collaboration):
                Phys.Lett. B {\bf 229}, 421 (1989).

\bibitem{IST}   Ishii, C., Saito, K. and Takagi, F.:
                Phys.Lett. B {\bf 216}, 409 (1989).

\bibitem{GRO}   Bosveld, G.D., Dieperink, A.E.L. and Scholten, O.:
                Phys.Lett. B {\bf 264}, 11 (1991);
%
                Scholten, O. and Bosveld, G.D.:
                Phys.Lett. B {\bf 265}, 35 (1991).

\bibitem{MTN}   Melnitchouk, W., Thomas, A.W. and Nikolaev, N.N.:
                Z.Phys. A {\bf 342}, 215 (1992).

\bibitem{MST}   Melnitchouk, W., Schreiber, A.W. and Thomas, A.W.:
                Phys.Rev. D {\bf 49}, 1183 (1994).

\bibitem{WEIN}  Weinberg, S.:
                Phys.Rev. {\bf 150}, 1313 (1966).

\bibitem{DLY}   Drell, S.D., Levy D.J. and Yan, T.M.:
                Phys.Rev. D {\bf 1}, 1035 (1970).

\bibitem{MTV}   Melnitchouk, W. and Thomas, A.W.,
                Phys.Rev. D {\bf 47}, 3794 (1993);
%
                Thomas, A.W. and Melnitchouk, W.,
                in: Proceedings of the JSPS-INS Spring School
                (Shimoda, Japan),
                (World Scientific, Singapore, 1993);
%
                Melnitchouk, W.,
                Ph.D. thesis, University of Adelaide, June 1993
                (unpublished).

\bibitem{ZOL}   Zoller, V.R.:
                Z.Phys. C {\bf 54}, 425 (1992);
%
                Holtmann, H., Szczurek, A. and Speth, J.:
                J\"ulich preprint KFA-IKP(TH) 1993-33.

\bibitem{AG}    Arakelyan, G. and A.Grigoryan, A.:
                Sov.J.Nucl.Phys. {\bf 34}, 745 (1981).

\bibitem{MSM}   Mulders, P.J., Schreiber, A.W. and Meyer, H.:
                Nucl.Phys. {\bf A549}, 498 (1992).

\bibitem{GCPP}  G\"uttner, F., Chanfray, G., Pirner, H.J. and Povh, P.:
                Nucl.Phys. {\bf A429}, 389 (1984).

\bibitem{PISF}  Shigetani, T., Suzuki, K. and Toki, H.:
                Phys.Lett. B {\bf 308}, 383 (1993).

\bibitem{SHAKIN} Shakin, C.M. and Sun, W.-D.:
                Phys.Rev. C {\bf 50}, 2553 (1994).

\bibitem{SMRS}  Sutton, P.J., Martin, A.D., Roberts, R.G.
                and Stirling, W.J.:
                Phys.Rev. D {\bf 45}, 2349 (1992).

\bibitem{NA10}  Betev, B., et al. (NA10 Collaboration):
                Z.Phys. C {\bf 28}, 15 (1985).

\bibitem{SB}    Brown, G.E., Buballa, M., Li, Z. and Wambach, J.:
                Stony Brook preprint SUNY-NTG-94-54;
%
                Buballa, M., preprint SUNY-NTG-94-61.

\bibitem{RS}    Rarita, W. and Schwinger, J.:
                Phys.Rev. {\bf 60}, 61 (1941).

\bibitem{BDM}   Benmerrouche, M., Davidson, R.M. and Mukhopadhyay, N.C.:
                Phys.Rev. C {\bf 39}, 2339 (1989).

\bibitem{FF}    Field, R.D. and Feynman, R.P.:
                Nucl.Phys. {\bf B136}, 1 (1978).

\bibitem{SSV}   Sloan, T., Smadja, G. and Voss, R.:
                Phys.Rep. {\bf 162}, 45 (1980).

\bibitem{SCHM}  Schmitz, N.:
                Int.J.Mod.Phys. A {\bf 3}, 1997 (1988).

\bibitem{GS}    Gehrmann, T and Stirling, W.J.:
                Z.Phys. C {\bf 65}, 461 (1995).

\bibitem{CTEQ}  Lai, H.L., et al. (CTEQ):
                preprint MSU-HEP-41024, hep-ph/9410404.

\bibitem{CK}    Carlitz, R. and Kaur, J.;
                Phys.Rev.Lett. {\bf 38} 674 (1977);
%
                Kaur, J.:
                Nucl.Phys. {\bf B128}, 219 (1977).

\bibitem{SCHAEFER} Sch\"afer, A.:
                Phys.Lett. B {\bf 208}, 175 (1988).

\bibitem{RW}    Renton, R. and Williams, W.S.C.:
                Ann.Rev.Nucl.Part.Sci. {\bf 31}, 193 (1981).

\bibitem{BRE}   Brenner, A.E., et al.:
                Phys.Rev. D {\bf 26}, 1497 (1982).

\bibitem{ACK}   Ackermann, H., et al.:
                Nucl.Phys. {\bf B120}, 365 (1977).

\bibitem{SLO}   Sukhatme, U.P., Lassila, K.E. and Orava, R.:
                Phys.Rev. D {\bf 25}, 2975 (1982).

\bibitem{BFM}   Bartl, A., Fraas, H., and Majoretto, W.:
                Phys.Rev. D {\bf 26}, 1061 (1982).

\bibitem{FFFRAG} Field, R.D. and Feynman, R.P.:
                Phys.Rev. D {\bf 15}, 2590 (1977).

\bibitem{LUND}  Andersson, B., Gustafson, G., Ingelman, G.
                and Sjostrand, T.:
                Phys.Rep. {\bf 97}, 31 (1983).

\bibitem{FPS}   Fontannaz, M., Pire, B. and Schiff, D.:
                Phys.Lett. {\bf 77} B, 315 (1978);
%
                Beavis, D. and Desai, B.R.:
                Phys.Rev. D {\bf 23}, 1967 (1981).

\bibitem{COUNT} Brodsky, S.J. and Blankenbecler, R.:
                Phys.Rev. D {\bf 10}, 2973 (1974);
%
                Brodsky, S.J. and Farrar, G.:
                Phys.Rev.Lett. {\bf 31}, 1193 (1975).

\bibitem{BFMS}  Bartl, A., Fraas, H., and Majoretto, W.:
                Z.Phys. C {\bf 6}, 335 (1980);
                {\em ibid} {\bf 9}, 181 (1981).

\bibitem{BIGI}  Bigi, I.I.Y.:
                Nuov.Cim. {\bf 41A}, 43, 581 (1977).

\bibitem{DON}   Donoghue, J.F.,
                Phys.Rev. D {\bf 17}, 2922 (1978);
                {\em ibid} D {\bf 19}, 2806 (1979).

\bibitem{HMC}   Mallot, G., et al.:
                Letter of intent, {\em Semi-inclusive muon scattering
                from a polarized target},
                preprint CERN/SPSLC 95-27 (March 1995).

\bibitem{AEK}   Alberg, M., Ellis, J., and Kharzeev, D.:
                Preprint CERN-TH/95-47 (February 1995).

\bibitem{LAMDAT} Arneodo, M., et al. (EM Collaboration):
                Phys.Lett. {\bf 145} B, 156 (1984);
%
                Hicks, R.G., et al.:
                Phys.Rev.Lett. {\bf 45}, 765 (1980);
%
                Cohen, I., et al.:
                Phys.Rev.Lett. {\bf 40}, 1614 (1978);
%
                Brock, R., et al.:
                Phys.Rev. D {\bf 25}, 1753 (1982).
\end{document}